# Ultrasound Beam Steering with Flattened Acoustic Metamaterial Luneburg Lens

**Liuxian Zhao[1], Eitan Laredo[2], Olivia Ryan[3], Amirhossein Yazdkhasti[4], Hyun-Tae Kim[4], Randy Ganye[4], Timothy Horiuchi[1,5], Miao Yu[1,4,*]**

[1]Institute for Systems Research, University of Maryland, College Park, MD, 20742, USA

[2]Department of Mechanical and Industrial Engineering, University of Massachusetts Amherst, Massachusetts 01002, USA

[3]Department of Electrical Engineering, Roger Williams University, Bristol, Rhode Island 02809, USA

[4]Department of Mechanical Engineering, University of Maryland, College Park, Maryland 20742, USA

[5]Department of Electrical and Computer Engineering, University of Maryland, College Park, Maryland 20742, USA

*Author to whom correspondence should be addressed: mmyu@umd.edu

**Keywords:** Flattened Acoustic Luneburg Lens, Quasi-conformal Transformation, Additive Manufacturing, Beam Steering

**ABSTRACT**

We report ultrasound beam steering based on 2D and 3D flattened acoustic metamaterial Luneburg lenses at 40 kHz. The effective properties of the lenses are obtained by using the quasi-conformal transformation (QCT) technique and solving the Laplace equation with Dirichlet and Neumann boundary conditions. A 2D lens and a 3D lens were designed and fabricated. The numerical and experimental results with these lenses demonstrate excellent beam steering performance of ultrasonic waves in both near field and far field.

A Luneburg lens is a type of gradient refractive index (GRIN) lens. Its index profile can be described as $n(r) = n_0\sqrt{2-(r/R)^2}$, where *r* is the distance from the lens center and *n₀* is the refractive index of the surrounding medium. For an ideal spherical Luneburg lens, an incident plane wave will be focused at the diametrically opposite point on its surface without aberration, and vice versa. While electromagnetic Luneburg lenses have been extensively studied for imaging and communication applications, acoustic Luneburg lenses have only recently attracted significant interest. Luneburg lenses based on phononic crystals and variable thickness structures have been designed for the manipulation of structural waves (e.g., [1, 2]). Recently, GRIN acoustic metamaterials have given rise to a number of novel acoustic devices, which have been explored for imaging, focusing, acoustic receivers, retroreflectors, and bi-functional devices [3-7]. Most of the current reported Luneburg lenses are 2D devices. A 3D acoustic Luneburg lens design has been attempted, but its numerical and experimental performance has not yet been reported [8].

One attractive application of an acoustic Luneburg lens is to obtain multi-directional beam control (e.g., beamforming or steering) without complex and expensive phase-shift networks required with conventional phase array techniques [9-12]. Beam steering can be realized by simply switching the feeding sources placed along the circular surface of the Luneburg lens. In comparison to the spherical (3-D) or rounded (2-D) surface of the standard Luneburg lens, a flattened Luneburg lens is more desirable for accommodating the signal sources in practical applications. Although multi-directional beam steering with a flattened Luneburg lens has been explored in the electromagnetic regime [13, 14], acoustic multi-directional beam steering with a flattened metamaterial lens has not been demonstrated.

Quasi-conformal Transformation (QCT) is an effective technique used for changing the structure geometry and refractive index without introducing anisotropy, which also helps reduce the computational complexity compared to traditional transformation optics [15-18]. This technique has been applied to produce a variety of electromagnetic devices with unique functions, such as the flattened dielectric lens, planar reflectors [19-21], and electromagnetic carpet cloaks [22, 23]. Recently, this technique has also been exploited to achieve acoustic devices, such as a planar acoustic metamaterial focusing lens [24], a broadband underwater acoustic carpet cloak [25], and a 2D flattened-Luneburg lens for focusing [26].

In this letter, we demonstrate beam steering in the ultrasound regime with 2D and 3D flattened acoustic metamaterial Luneburg lenses. When a point source is moved along the flat surface of the 2D and 3D lenses, the outgoing plane wave is steered in different directions on the 2D plane (Figure 1(a)) or in 3D space (Figure 1(b)), respectively. The lens design is based on QCT and the quasi-orthogonal meshes are obtained by solving the Laplace's equation with Dirichlet and Neumann boundary conditions, instead of using commercial software meshing tools. Simulation and experimental results show that the 2D and 3D lenses can effectively achieve beam steering when a feeding source is placed at different positions along the flattened lens surface for the designed working frequency of 40 kHz.

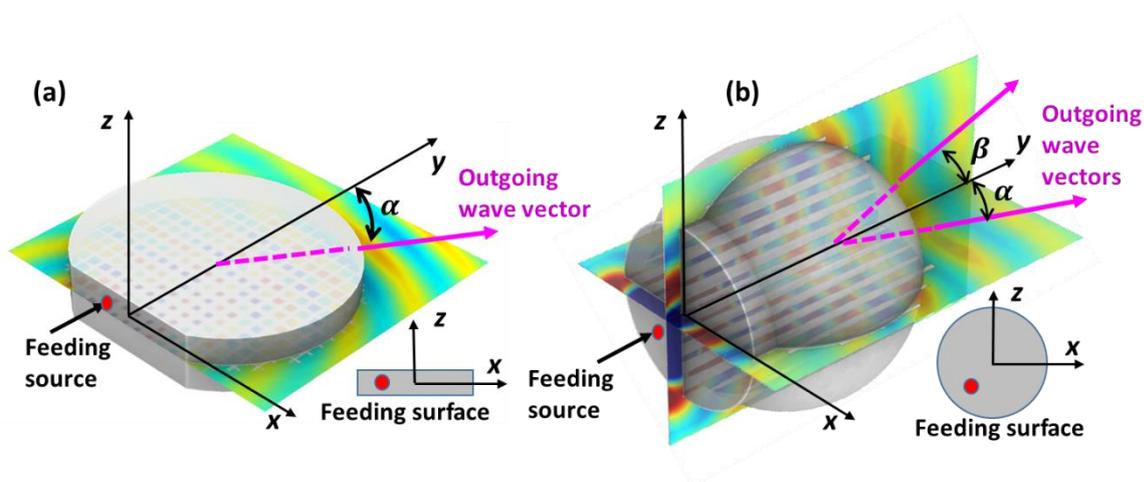

**Figure 1: Ultrasound beam steering using the flattened acoustic Luneburg lens. (a) 2D**

**beam steering. When the point source is moved along *x* axis on the flattened feeding surface, the outgoing plane wave is steered in the horizontal (*xy*) plane. The beam direction can be characterized by the angle *α* with respect to *y* axis. (b) 3D beam steering. When the point source is moved in the *xz* plane on the flattened feeding surface, the outgoing plane wave is steered in a 3D space. The beam direction can be characterized by two angles: *α* in the horizontal (*xy*) plane with respect to *y* axis and *β* in the vertical (*yz*) plane with respect to *y* axis. The insets show representative locations of the source on the feeding surfaces.**

To achieve a flattened acoustic Luneburg lens, we employ the QCT method described by Chang *et al*. [27]. This approach can help achieve material parameter transformation with a minimum amount of anisotropy by solving Laplace equations with proper boundary conditions. The specific geometries used for our QCT are shown in Figure 2 (a). Here the original circular acoustic Luneburg lens, denoted as the virtual space in Figure 2(a), is transformed into a flattened acoustic Luneburg lens denoted as the physical space in Figure 2(b). The distribution of flattened Luneburg lens in the physical space is calculated via solving Laplace's equation with Dirichlet and Neumann boundary conditions as given below:

$$\begin{cases} AB|_x = EF|_x = x' \\ AF|_y = BC|_y = CD|_y = DE|_y = y' \\ \vec{n} \cdot \nabla_x|_{BC,CD,DE,AF} = 0 \\ \vec{n} \cdot \nabla_y|_{AB,EF} = 0 \end{cases}, \qquad (1)$$

where (*x*, *y*) are the coordinates of the virtual space and (*x'*, *y'*) are the coordinates of the physical space, $\vec{n}$ represents the outward normal vector to the surface boundaries, and $\nabla$ is the gradient operator.

We then obtain an effective refractive index for the flattened acoustic Luneburg lens $n'$ based on the equation [24]:

$$n'^2 = \frac{n^2}{\det(J)n_0^2} , \qquad (2)$$

where $n$ indicates the refractive index of original Luneburg lens and $J$ is the Jacobian matrix expressed as:

$$J = \begin{bmatrix} \frac{\partial x'}{\partial x} & \frac{\partial x'}{\partial y} \\ \frac{\partial y'}{\partial x} & \frac{\partial y'}{\partial y} \end{bmatrix} . \qquad (3)$$

Because of the quasi-orthogonal coordinates employed in both spaces, we have the following approximation conditions: $\frac{\partial x'}{\partial y} \approx 0$ and $\frac{\partial y'}{\partial x} \approx 0$.

After obtaining an in-plane 2D pattern of the refractive index based on Eqs. (2) and (3), the refractive index pattern is extended along the thickness direction to obtain the 2D lens. Furthermore, a 3D pattern of refractive index is obtained by rotating the 2D index along its *y*-axis. In this work, we choose a radius of acoustic Luneburg lens is $R = 2$ cm and an open angle of $\varphi = 90°$, which can achieve beam steering in the angular range of -45º to 45º. COMSOL, a commercially-available software package is used to solve the Laplace equation with the corresponding boundary conditions. The refractive-index distributions normalized by the background air ($n/n_0$ and $n'/n_0$) and the quasi-orthogonal meshes are shown in Figure 2(a) and (b) for the virtual space and the physical space, respectively.

According to Figure 2(b), the refractive index range of the flattened acoustic Luneburg lens is between 1 and 1.82, which can be achieved by using acoustic metamaterials with varying unit cell structures. Here, we use a 3D lattice truss unit cell [8, 28], as shown in Figure 2(c). It consists of three orthogonal beams, so that each cell is interconnected with its adjacent cells to form a self-supporting lattice. In our design, we choose the unit cell geometric parameters $a = b = a_0 \times d$. Here, $d$ is the periodicity of the unit cell, which is chosen to be 2 mm. In principle, a range of refractive index values can be obtained by changing the unit cell geometric factor $a_0$.

The refractive index of each unit cell is calculated based on a standard retrieval method [29]. Note that the refractive index is a function of frequency. For example, for $a_0 = 0.6$ in Figure S1(a) of the Supplementary Material, the frequency dependence of refractive index is clearly shown. In this study, the ultrasonic frequency $f = 40$ kHz is selected for investigation. Other acoustic frequency regimes or broadband operations can be achieved through carefully choosing the design parameters. At 40 kHz, the selected refractive indices $n'/n_0$ ( from 1.07 to 1.82) for achieving the flattened Luneburg lens can be obtained by with geometric factor $a_0$ that range from 0.23 to 0.77, as shown Figure S1(b) in the Supplementary Material. In this work, two flattened acoustic Luneburg lenses (2D and 3D devices) were designed by stacking a series of 3D truss unit cells layer-by-layer to form a stable lattice, as shown in Figure 2(d) and (e).

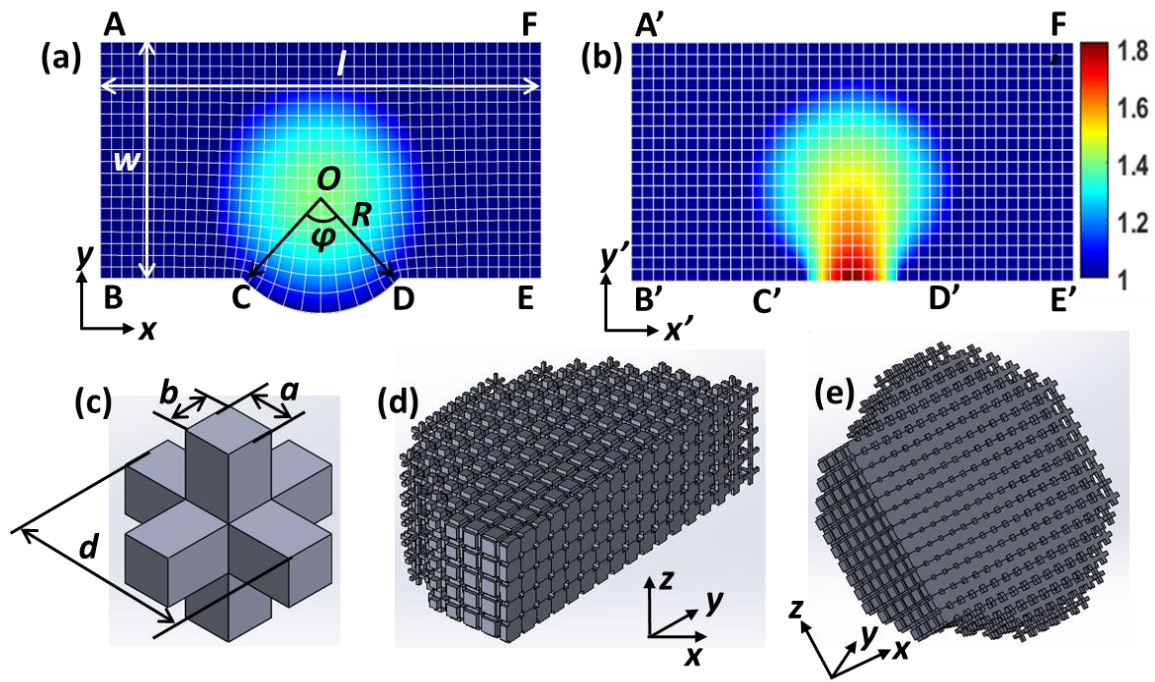

**Figure 2: Design of the flattened acoustic Luneburg lens. (a) and (b) Refractive index patterns for the circular acoustic Luneburg lens in the virtual space and flattened acoustic Luneburg lens in physical space. The dimension of the entire space is $l \times w = 4R \times 2R$. (c)**

**An example of the truss unit cell. (d) Cross-section views of 2D lens with five stacking layers along the *z* direction and (e) 3D lens with 19 stacking layers along the *z* direction.**

The 2D and 3D flattened, acoustic Luneburg lens designs (Figure 3(a) and (b)) were fabricated via 3D printing (Stratasys Objet500 Connex3 3D printer) with a maximum resolution of 16 µm. The experimental setup used for the device characterization in an anechoic chamber is shown in Figure 3(c) and (d). A compact speaker with diameter of 9 mm (MA40S4S from Murata Manufacturing Co., Ltd.) driven by a pre-amplifier (Sony STR-DH100) was used to generate an approximate point source for excitation. A fiber optic probe [30] was used to measure the acoustic field distribution. A condenser microphone (Type 4138 from Bruel & Kjaer) was used as a reference sensor to obtain the phase distribution. A rotational stage, a radial translational stage, and a vertical translational stage were used to move the fiber optic probe along the circumferential, radial (*y*) and vertical (*z*) directions to obtain 2D and 3D acoustic field measurements. Acoustic pulse signals with a central frequency of $f = 40$ kHz were generated by the speaker using a signal amplitude of $V_p = 5$ V.

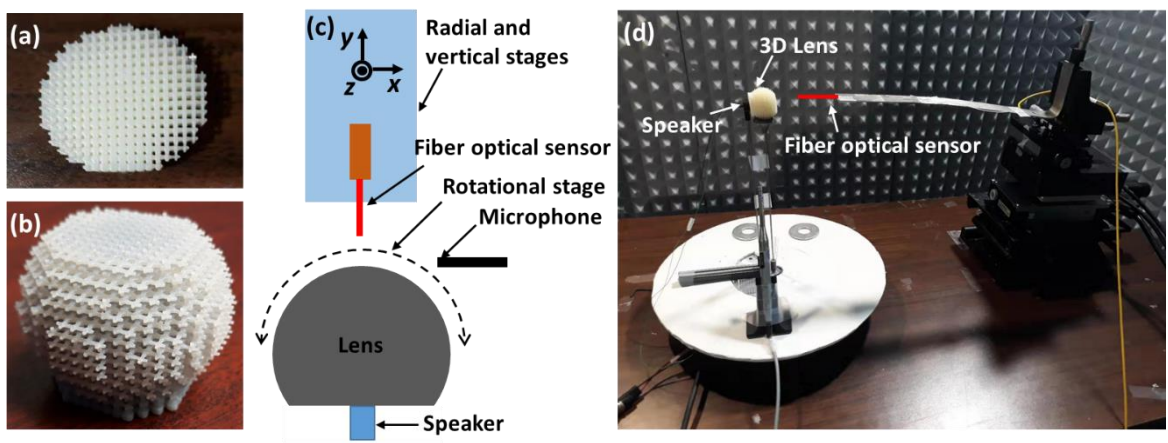

**Figure 3: Experimental characterization of full-field wave propagation in the near field. Photos of (a) 2D and (b) 3D flattened acoustic metamaterial Luneburg lenses. (c) Schematic and (d) photo of the experimental setup.**

We first characterized the acoustic beam steering capability of the 2D flattened acoustic Luneburg lens. When a speaker is moved along the central line of the flat side of the lens, the azimuthal angle $\alpha$ of the outgoing acoustic wave is expected to steer from -45º to 45º. For simplicity without losing the generality, we chose four speaker locations with a separation of 2 mm, which corresponds to steering angles $\alpha$ of approximately 0º, 10º, 20º, and 30º. At each feeding source location, full 3D acoustic wave simulations were performed to obtain the pressure waveforms (Figure 4(a)) and amplitudes (Figure 4 (c)) with a 40 kHz point source excitation. The simulations were conducted by using the commercial software package COMSOL. Radiation boundary conditions were applied on the outer boundaries to assume infinite air spaces. The simulation results were collected from the central plane ($z = 0$) of the five layers.

In the physical experiments, the 2D lens was characterized with the excitation at the four speaker locations by moving the speaker along the flat side (with 2 mm steps). At each speaker location, full spatial measurements of the pressure waveforms (Figure 4(b)) and amplitudes (Figure 4(d)) were obtained in the central plane of the lens for radial distances from 24 mm to 60 mm with a step size of 2 mm. Both the simulation and experimental results show that the flattened 2D Luneburg lens is capable of converting acoustic waves generated from a point source (or a small speaker) into plane waves. By changing the feeding source locations, the main acoustic wave energy is clearly steered to the expected directions. The experimental results are shown to be in good agreement with the simulation results.

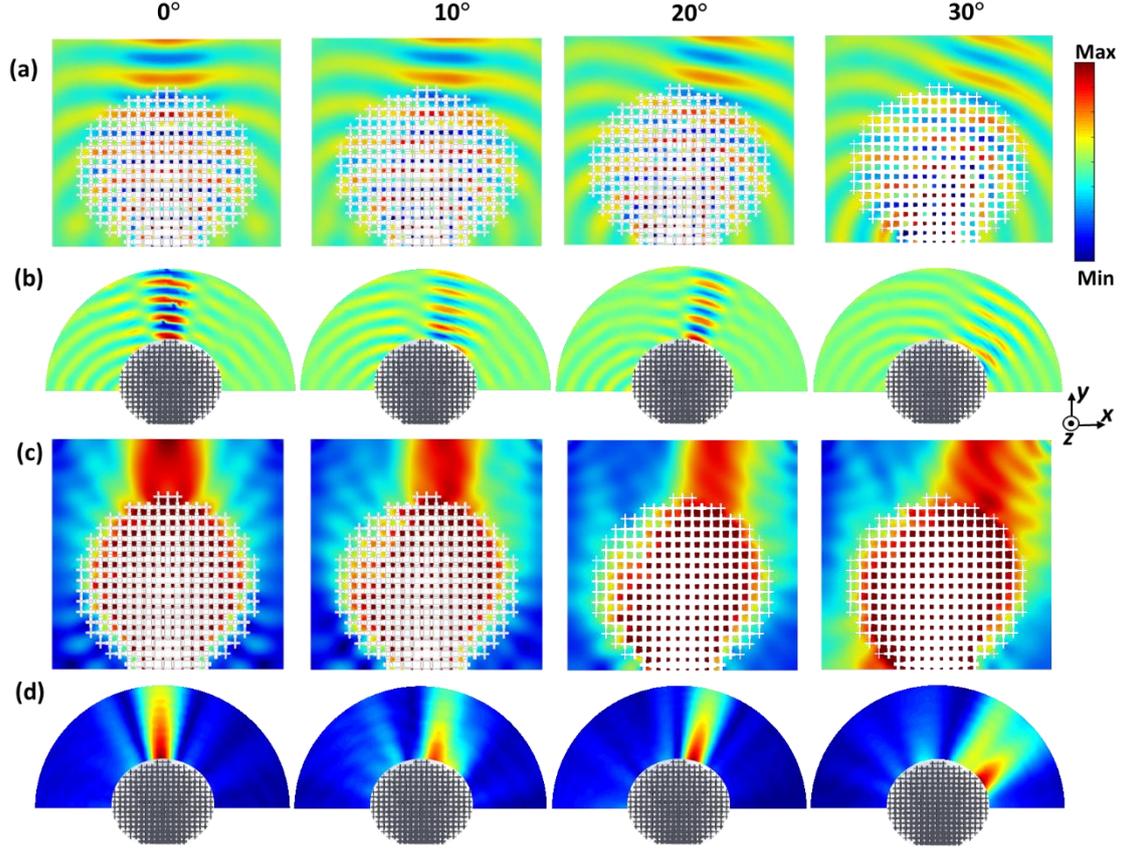

**Figure 4: Simulation results (a) and (c) and experimental results (b) and (d) of beamforming with 2D flattened acoustic metamaterial Luneburg lens. The point source excitation is moved to the four feeding locations of *x* = 0 mm, -2 mm, -4 mm, and -6 mm (*y*=0 on the feeding surface) to achieve 0°, 10°, 20°, and 30° steering angles, respectively. (a) and (b) are waveform distributions, and (c) and (d) are pressure amplitude distributions. All the results are obtained in the middle layer (*z*=0) of the 2D lens.**

Next, we investigated the 3D flattened Luneburg lens for beaming steering in a 3D space. Since the 3D lens is circularly symmetric around the *y* axis, its beam steering performance was characterized for the source located at (*x*= -6, *z*=0) on the feeding surface. By using the same approach as described for the 2D lens, the 3D simulation results (Figure 5(a)-(c)) and experimental waveform measurement results (Figure 5(d) and (e)) in two planes (horizontal *xy* plane and vertical *yz* plane) were obtained with the 3D lens. These results successfully demonstrate the 3D beam steering performance of the lens. The propagation direction of the outgoing plane wave can be described by using two angles: $\alpha$ in the horizontal plane and $\beta$ in the vertical plane (see Figure 1 (b)). For the selected feeding source location ($x=$ -6, $y=0$, $z=0$), the beam is shown to be steered to a direction with $\alpha= 30°$ (Figure 5 (b) and (d)) in the horizontal plane and $\beta =0°$ (Figure 5 (c) and (e)) in the vertical plane. The experimentally-measured waveform patterns (Figure 5 (d) and (e) exhibit close agreement with the simulation results (Figure 5 (b) and (c)) in the selected planes.

In addition, the far-field acoustic radiation was measured experimentally for both 2D and 3D lenses (see Figure S2 in the supplementary material), which clearly demonstrate that different steering angles can be achieved in the far field by changing the feeding source locations. It is worth noting that the 3D lens demonstrates excellent far-field 3D beam steering performance for the tested 5 different feeding source locations. Overall, both 2D and 3D lenses exhibit good performance for the designed beam steering range of -45º to 45º.

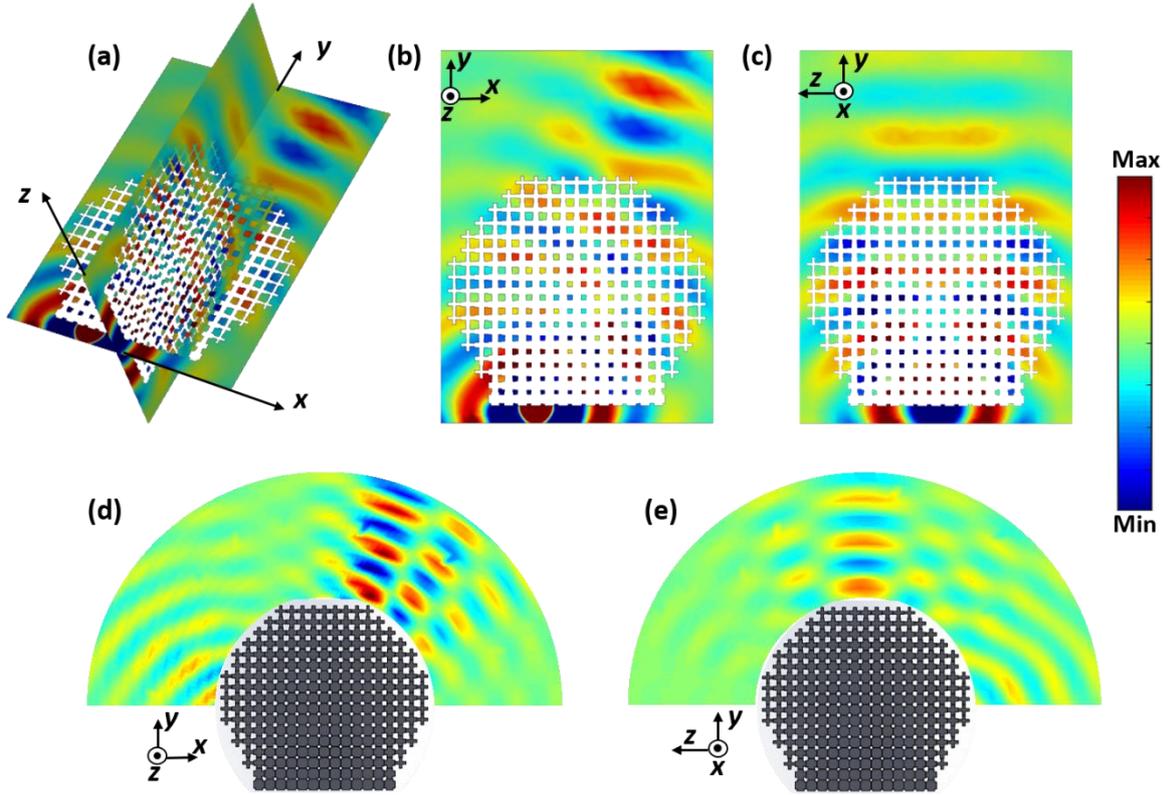

**Figure 5: Waveform patterns of beam steering with 3D flattened acoustic metamaterial Luneburg lens.** (a)-(c) simulation results showing 3D simulations of both horizontal plane and vertical plane, horizontal plane only, and vertical plane only, respectively. (d) and (e) experimental results obtained from horizontal plane and vertical plane, respectively. The acoustic source is located at ($x=$ -6, $z=0$) on the feeding plane $xz$ ($y=0$).

We have demonstrated the design and performance characterization of both 2D and 3D flattened acoustic Luneburg lenses for ultrasound beam steering. The Luneburg lenses are based on GRIN acoustic metamaterials, and the QCT technique with Dirichlet and Neumann boundary conditions is used to achieve the flattened lenses. The beam steering performance of both the 2D and 3D lenses were experimentally demonstrated, which compare well with the numerical simulations. Although these lenses were designed to operate at 40 kHz, the present work can be extended to other acoustic frequency regimes or broadband operations. These flattened acoustic Luneburg lenses hold great promise for applications including ultrasonic imaging, diagnosis, treatment, and sonar systems.

## Acknowledgements

This work was supported by the AFOSR Center of Excellence on Nature-Inspired Flight Technologies and Ideas and NSF (CMMI1436347).

## Conflict of Interest

The authors declare no conflict of interest.